\documentclass[twocolumn,10pt,prb,aps,showpacs,superscriptaddress,preprintnumbers,amsmath,amssymb]{revtex4-1}

\usepackage{graphicx} 
\usepackage{dcolumn} 
\usepackage{bm} 

\usepackage{ulem}
\usepackage[]{color}

\def \vecq {\textbf{q}}
\begin{document}

\title{The quartet ground state in CeB$_6$: an inelastic x-ray scattering study}

\author{M.~Sundermann}
  \affiliation{Institute of Physics II, University of Cologne, Z{\"u}lpicher Stra{\ss}e 77, 50937 Cologne, Germany}
  \affiliation{Max-Planck-Institute for Chemical Physics of Solids, N{\"o}thnizer Stra{\ss}e 40, 01187 Dresden, Germany}
\author{K.~Chen}
  \affiliation{Institute of Physics II, University of Cologne, Z{\"u}lpicher Stra{\ss}e 77, 50937 Cologne, Germany}
\author{H.~Yava\c{s}}
  \affiliation{PETRA III, Deutsches Elektronen-Synchrotron (DESY), Notkestra{\ss}e 85, 22607 Hamburg, Germany}
\author{Han-Oh Lee}
  \affiliation{Center for Correlated Matter, Zhejiang University, Hangzhou 310058, People's Republic of China}
\author{Z.~Fisk}
  \affiliation{Department of Physics and Astronomy, University of California, Irvine, CA 92697, USA}
\author{M.~W.~Haverkort}
  \affiliation{Max-Planck-Institute for Chemical Physics of Solids, N{\"o}thnizer Stra{\ss}e 40, 01187 Dresden, Germany}
  \affiliation{Institute for Theoretical Physics, Heidelberg University, Philosophenweg 19, 69120 Heidelberg, Germany}
\author{L.~H.~Tjeng}
  \affiliation{Max-Planck-Institute for Chemical Physics of Solids, N{\"o}thnizer Stra{\ss}e 40, 01187 Dresden, Germany}
\author{A.~Severing}
  \affiliation{Institute of Physics II, University of Cologne, Z{\"u}lpicher Stra{\ss}e 77, 50937 Cologne, Germany}
  \affiliation{Max-Planck-Institute for Chemical Physics of Solids, N{\"o}thnizer Stra{\ss}e 40, 01187 Dresden, Germany}

\begin{abstract}
We investigated the ground state symmetry of the cubic hidden order compound CeB$_6$ by means of core level non-resonant inelastic x-ray scattering (NIXS). The information is obtained from the directional dependence of the scattering function that arises from higher than dipole transitions. Our new method confirms that the ground state is well described using a localized crystal-field model assuming a $\Gamma_8$ quartet ground state.
\end{abstract}

\maketitle

\section{Introduction}
\begin{figure}[]
	\centering
\includegraphics[width=0.8\columnwidth]{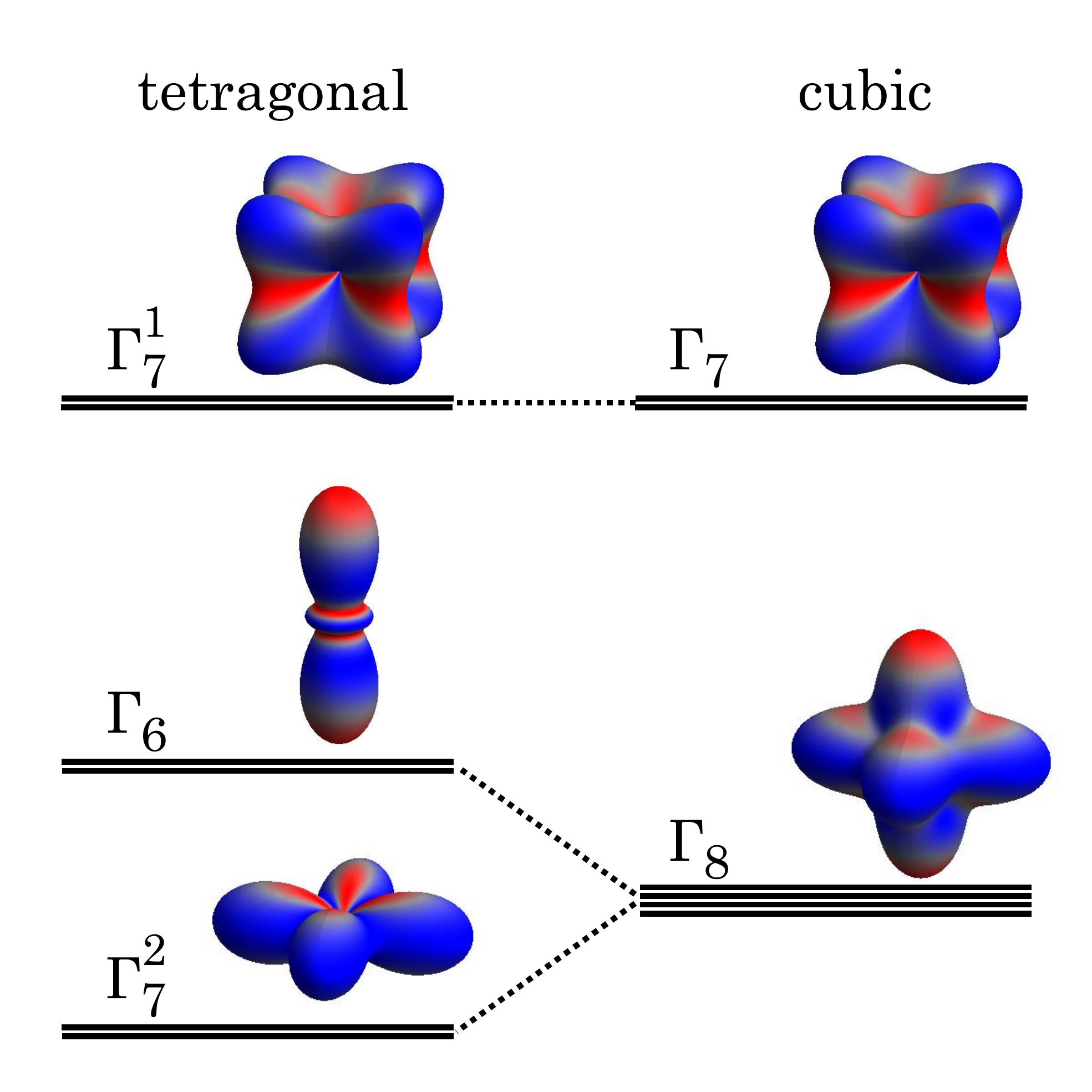}
\caption{Electron density plots for an $f$-electron in Ce3$^+$, left for tetragonal and right for cubic point symmetry. For tetragonal symmetry the crystal-field states expressed in $J_z$ representation are  
$\Gamma_6$\,=\,$|\pm1/2 \rangle$, 
$\Gamma_7^1$\,=\,$\alpha$$|\pm5/2\rangle\,-\,\sqrt{1-\alpha^2}$$|\mp\,3/2 \rangle$ and 
$\Gamma_7^2$\,=\,$\sqrt{1-\alpha^2}$$|\pm5/2\rangle$\,+\,$\alpha$$|\mp 3/2 \rangle$ with 
$\alpha^2$\,$\le$\,1; for cubic symmetry $\alpha\,=\,\sqrt{1/6}$ so that 
$\Gamma_7^1$\,=\,$\sqrt{1/6}$$|\pm5/2\rangle\,-\,\sqrt{5/6}$$|\mp\,3/2 \rangle$ and 
$\Gamma_8$\,=\,($|\pm1/2 \rangle$;$\sqrt{5/6}$$|\pm5/2\rangle$\,$+$\,$\sqrt{1/6}$$|\mp 3/2 \rangle$).}
\label{Fig01}
\end{figure}

The material class of rare earth hexaborides has attracted considerable attention over the years. It comprises of a variety of different fascinating ground states (see Ref. \cite{Sun2016} and references therein) which include exotic magnetically ordered phases, heavy fermion behavior, as well as Kondo insulating ground states. CeB$_6$ is an important member of this material class, well known for its so-called hidden magnetic order. The very recent theoretical suggestion that SmB$_6$ could be a strongly correlated topological insulator \cite{Dzero2010,Takimoto_2011} even caused a flurry of new investigations (see Ref. \cite{Dzero2016} and references therein), thereby raising speculations that also YbB$_6$ under pressure could be topological. The standard and at the same time pressing question in all these studies concerns the symmetry of the ground state wave function of the crystal-electric field split 4$f$ multiplet. Here we explore the feasibility of using a recently developed experimental method, namely core-level non-resonant x-ray scattering (NIXS), to determine the ground state wave function of  CeB$_6$, a system which crystallizes in the cubic CsCl structure. Fig.\,\ref{Fig01} displays how the crystal-electric field splits the sixfold degenerate $j=5/2$ multiplet state of the Ce\,$4f^1$ into a $\Gamma_8$ quartet and $\Gamma_7$ doublet. 

CeB$_6$ is a heavy fermion compound that has been intensively studied for its rich magnetic phase diagram \cite{Effa1985}. Upon cooling CeB$_6$ enters a hidden order phase at 3.2\,K followed by an antiferromagnetic phase below 2.4\,K. The hidden order parameter is not accessible with e.g.\ neutron or standard x-ray diffraction at zero field. The application of an external field, however, induces  a dipole component with the wave vector of the quadrupolar ordering \cite{Erekelens1987}. Theory suggests that the multipolar moments of the localized 4$f$ electrons interact with each other via the itinerant 5$d$ conduction electrons, breaking up the fourfold ground state degeneracy of the Ce\,4$f$ wave function in the cubic crystal field stabilizing an antiferro-quadrupolar (AFQ) order \cite{Thalmeier1997}, a conjecture that by now has received credibility from a resonant x-ray diffraction study \cite{Matsumura2009,Lovesey2002}. The observation of a spin resonance in the inelastic neutron data of CeB$_6$ \cite{Friemel2012,Portnichenko2016} shows the importance of itinerancy for the formation of the multipolar and magnetic order \cite{Thalmeier2012}, the latter being supported by electronic structure investigations of CeB$_6$ \cite{Neupane2015,Koitzsch2016}. Inelastic neutron scattering finds in agreement with Raman scattering a crystal-field excitation at 46\,meV and it is generally accepted that the intriguing magnetic properties of CeB$_6$ evolve out of the fourfold degenerate $\Gamma_8$ ground state. The quartet ground state had been originally deduced from an unusual low temperature shift of the crystal-field excitation in Raman and inelastic neutron scattering data \cite{Zirn1984,Loew1985}. The energy shift was interpreted as a splitting of the quartet ground state in the low temperature phase in accordance with electron paramagnetic resonance (EPR) measurements\,\cite{Terzioglu2001}. A quartet ground state is also consistent with findings of the magnetic anisotropy \cite{Sato1984} and magnetic neutron form factor measurements\,\cite{Givord2003}. We have now revisited the symmetry aspect of CeB$_6$ in the paramagnetic phase using core level Ce\,N$_{4,5}$ ($4d\,\rightarrow\,4f$) NIXS, a  spectroscopic technique that directly probes the charge distribution of the Ce\,$4f$ electrons. 

\section{Spectroscopic Technique}

In the recent past we have shown that soft x-ray absorption spectroscopy (XAS) with linear polarized light is a very useful local probe for determining the anisotropy of wave functions in tetragonal \cite{HansmannPRL100} or orthorhombic heavy fermion compounds \cite{Strigari2012}, and for detecting small variations with unprecedented accuracy \cite{WillersPNAS2015}. However, for cubic compounds XAS cannot be applied since it relies on dipole transitions which cannot distinguish between the $\Gamma_8$ quartet and $\Gamma_7$ doublet state (see Fig.\,\ref{Fig01}). We have therefore performed an experiment that probes the symmetry with higher multipole transitions. This can be realized in a core level non-resonant inelastic x-ray scattering (NIXS) experiment with large momentum transfers $|\vecq{}|$. For large enough $|\vecq{}|$ the expansion of the transition operator e$^{i\vecq{}\textbf{r}}$ in the scattering function S($\vecq{}$,$\omega$) can no longer be truncated after the first term and as a result higher multipole terms contribute to S($\vecq{}$,$\omega$). These extra multipole contributions then give information that is not accessible in a dipole experiment \cite{LarsonPRL99,HaverkortPRL99,Gordon2008,GordonProceedings2009,BradleyPRB81,CaciuffoPRB81,SenGuptaPRB84,BradleyPRB84,GordonJElecSpec184,HiraokaEPL96,vanderLaanPRL108}. 

Bradley \textit{et al.} \cite{BradleyPRB84} and Gordon \textit{et al.} \cite{GordonJElecSpec184} were the first to observe higher multipole transitions in rare earth materials at the N$_{4,5}$ core level excitation for large momentum transfers $|\vecq{}|$ and the data were well described with a local many body approach by Haverkort \textit{et al.} \cite{HaverkortPRL99}. Already the early papers suggested that vector $\vecq{}$ dependent NIXS experiments on a single crystal should give insight into the ground state symmetry in analogy to an XAS experiment with linear polarized light \cite{HaverkortPRL99,Gordon2008,GordonJElecSpec184,BradleyPRB84}, and indeed, an experiment on cubic single crystals of MnO and CeO$_2$ at the Mn\,M$_{2,3}$ and Ce\,N$_{4,5}$ edges revealed direction dependencies in the higher multipole scattering function \cite{GordonProceedings2009}. Very recently, NIXS has been successfully used to determine the ground state symmetry and/or determine the rotation of the $f$ orbitals in fourfold symmetry in Ce single crystals \cite{WillersPRL109,Rueff2015,Sundermann2015}.

\begin{figure}[]
	\centering
	\includegraphics[width=1.0\columnwidth]{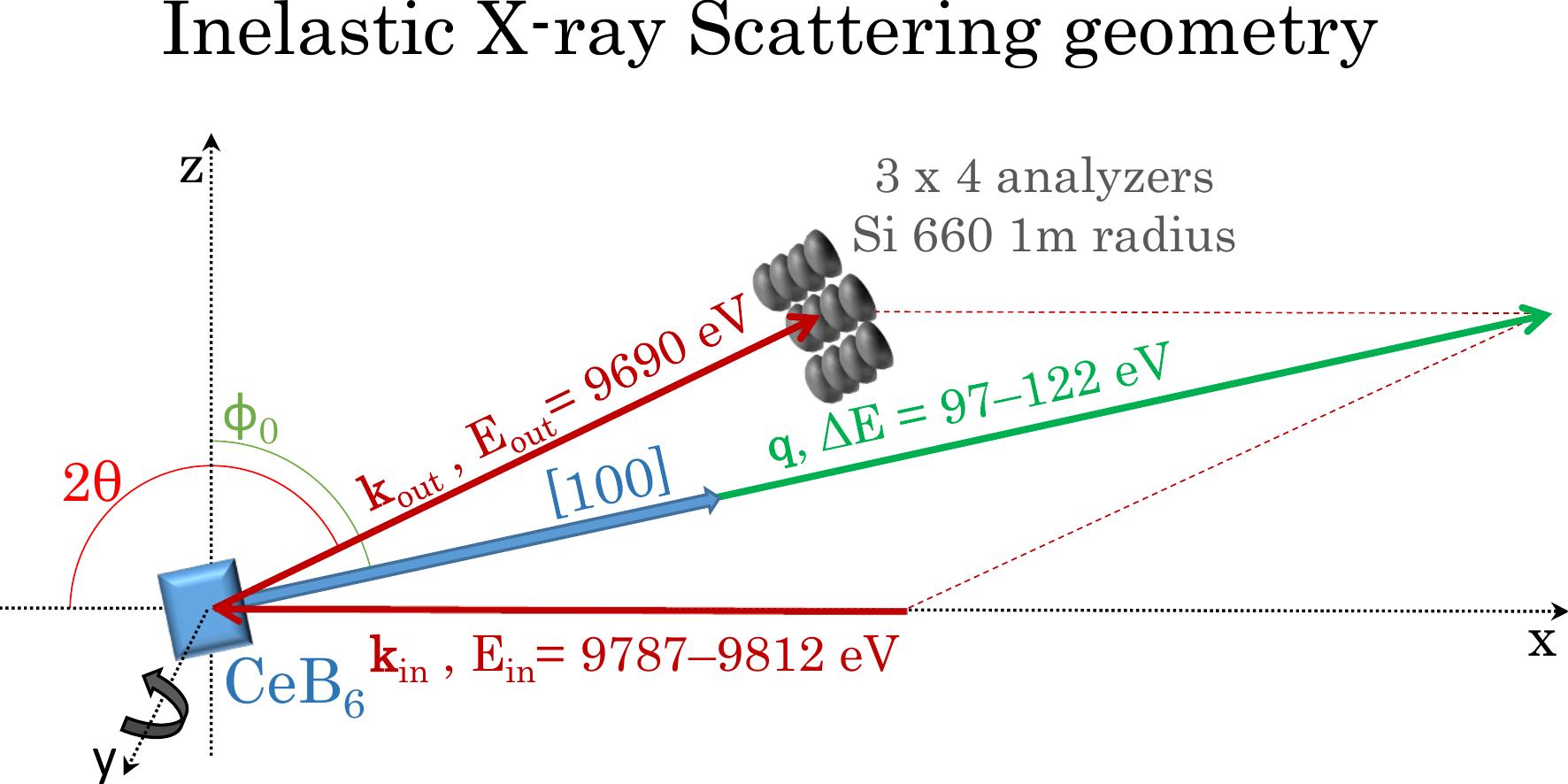}
	\caption{Scattering geometry of the NIXS experiment for a typical N$_{4,5}$ edge scan as in Fig.\,04, here for a scan with $\vecq{}$\,$\|$\,[100].}
\label{Fig02}
\end{figure}

\section{Experimental}

The single-crystal samples of CeB$_6$ were grown by the Al-flux method. Typically 0.7g of CeB$_6$ (as the elements) are heated with 60\,g of high purity Al (59) to 1450\,C, held there for 8\,hr and then cooled to 1000\,C at 2\,C/hr, when the furnace is shut off. The crystals are leached from the Al in NaOH solution.

The NIXS measurements on the CeB$_6$ Ce\,N$_{4,5}$ core level were performed at the beamline P01 of PETRA-III. The incident energy was selected with a Si(311) double monochromator. The P01 NIXS end station has a vertical geometry with twelve Si(660) 1\,m radius spherically bent crystal analyzers that are arranged in 3 x 4 array (see Fig.\,\ref{Fig02}). The fixed final energy was 9690\,eV. The analyzers were positioned at scattering angles of 2\,$\theta$\,$\approx$\,150$^\circ$, 155$^\circ$, and 160$^\circ$ which corresponds at elastic scattering to an averaged momentum transfer of $|\vecq{}|$\,=\,(9.6\,$\pm$\,0.1)\,\AA$^{-1}$. The scattered beam was detected by a position sensitive custom-made Lambda detector, based on a Medipix3 chip detector. The elastic line was regularly measured and pixelwise calibration yields an instrumental energy resolution of FWHM\,$\approx$\,0.7\,eV. A sketch of the scattering geometry, showing the incoming and outgoing photons as well as the transferred momentum $|\vecq{}|$, is given in Fig.\,\ref{Fig02} for a scan with $\vecq{}$\,$\|$\,[100] in specular geometry. In order to realize another crystallographic direction, e.g.\ $\vecq{}$\,$\|$\,[110], the sample can be turned with respect to the scattering triangle, or a different sample with another polished surface may be mounted in specular geometry.

Two crystals with (100) and (110) surfaces were mounted in a vacuum cryostat with Kapton windows. The measurements were performed with a pressure in the 10$^{-6}$\,mbar range. The two samples were oriented such that for $\vecq{}$\,$\|$\,[100] and $\vecq{}$\,$\|$\,[110] a specular scattering geometry was realized,\,i.e. with the surface normal parallel to the momentum transfer ($\phi$\,=$\phi_o$\,=\,$\theta$).  In order to check reliability, the $\vecq{}$\,$\|$\,[110] measurement was repeated on the (100) crystal but with the surface normal being rotated 45$^\circ$ away from $\vecq{}$ ($\phi$\,=\,$\phi$$_\circ$\,-\,45$^\circ$). The data were fully consistent. The $\vecq{}$\,$\|$\,[111] situation was realized by turning the (110) crystal to $\phi$\,=\,$\phi$$_\circ$\,-\,35$^\circ$.

\section{Results and Discussion}

\begin{figure}[]
	\centering
	\includegraphics[width=1.0\columnwidth]{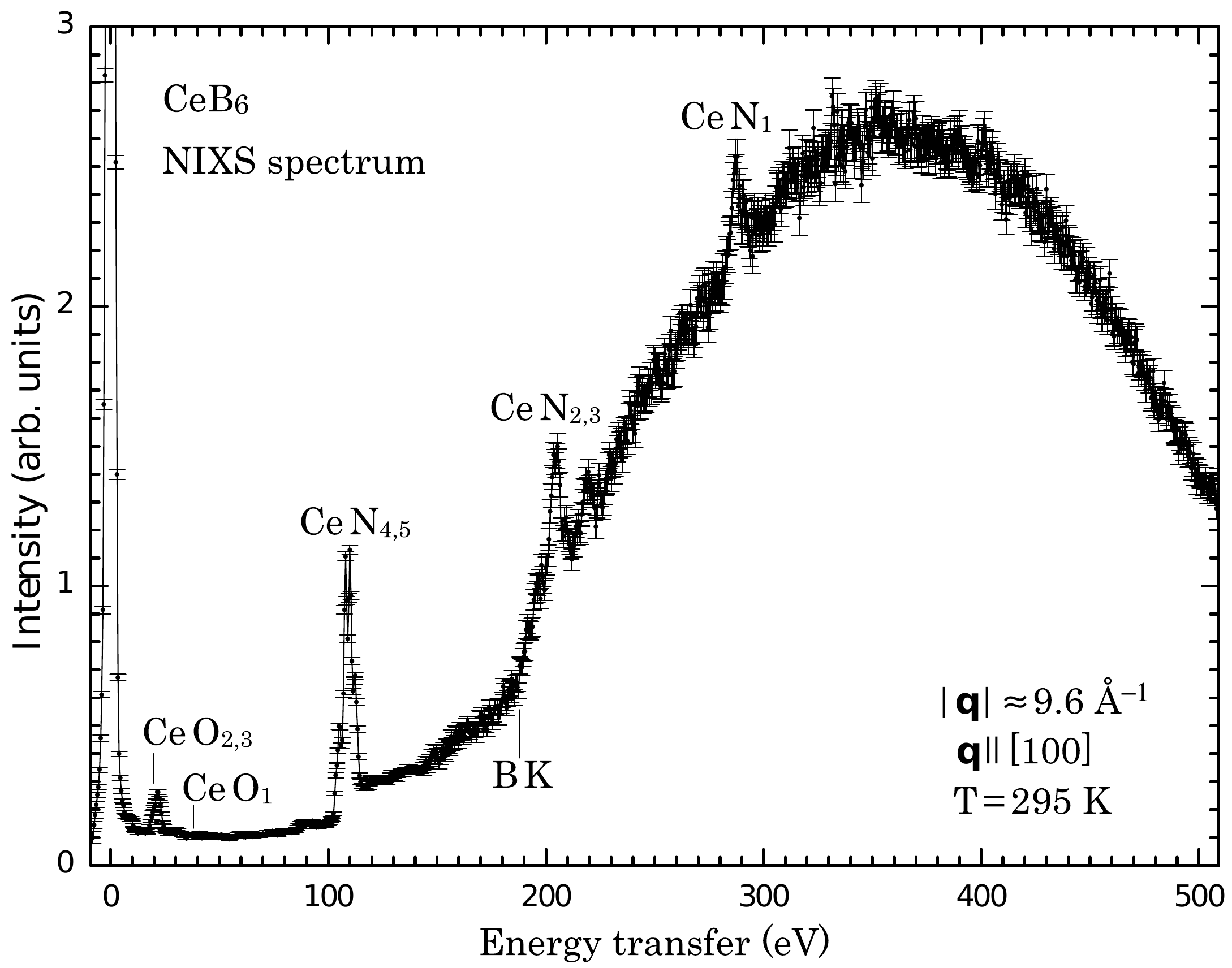}
	\caption{Experimental NIXS spectra of CeB$_6$: a wide scan covering the Ce\,O$_5$, N$_{4,5}$, N$_{2,3}$, and N$_{1}$ edges, the B\,K edge as well as the Compton signal. The direction of momentum transfer is $\vecq{}$\,$\|$\,[100]. }
\label{Fig03}
\end{figure}

Fig.\,\ref{Fig03} shows the NIXS spectrum across  the Ce\,N$_{4,5}$ ($4d\,\rightarrow\,4f$), N$_{2,3}$ ($4p\,\rightarrow\,4f$), and N$_{1}$ ($4s\,\rightarrow\,4f$) edges. The accompanying Compton contribution has its maximum at about 350 eV energy transfer. It is important to note that the Ce white lines are clearly discerned from the Compton scattering, and that especially the Ce\,N$_{4,5}$ white lines stand out with an excellent signal to background ratio. This shows that N$_{4,5}$ NIXS is an extremely suitable experimental method for the study of the local electronic structure of CeB$_6$, and for that matter, the class of rare-earth hexaborides. 

\begin{figure}[]
	\centering
	\includegraphics[width=1.0\columnwidth]{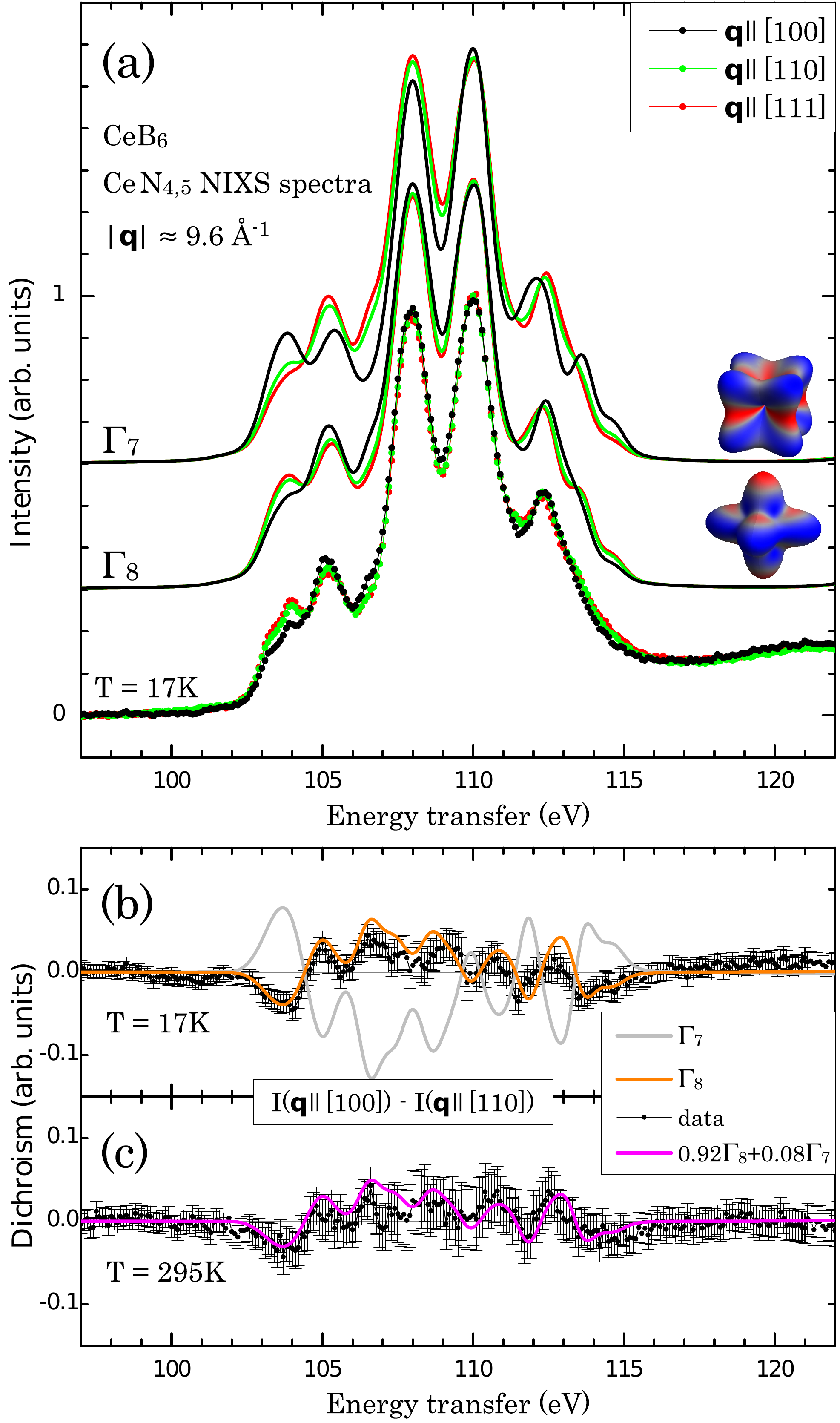}
	\caption{Top panel (a): calculated and experimental NIXS spectra of the Ce\,N$_{4,5}$-edge for the three transferred momentum directions $\vecq{}$\,$\|$\,[100], [110], and [111]. Bottom panel: difference spectra I($\vecq{}$\,$\|$\,[100])-I($\vecq{}$\,$\|$\,[110]) (black dots) (b) at low T and (c) at room temperature and respective simulations (see text).}
\label{Fig04}
\end{figure}

The top panel(a) of Fig.\,\ref{Fig04} shows the Ce\,N$_{4,5}$ NIXS spectra of CeB$_6$ (dots) taken at 17\,K, for the three momentum directions $\vecq{}$\,$\|$\,[100] (black dots),$\|$\,[110] (green dots), and $\|$\,[111] (red dots). The temperature of 17\,K is low enough to assure that only the ground state is populated. We recall that the excited crystal-field state is 46\,meV above the ground state \cite{Zirn1984,Loew1985}. Here only a constant background has been subtracted to account for the (weak) Compton signal (about 12\% of the signal peak) (see Fig.\,\ref{Fig03}). The size of the dots resembles the statistical error bar. 

There is a clear direction dependence that shows up strongest in the energy interval of 103 to 106\,eV. Especially the $\vecq{}$\,$\|$ [100] direction differs from the $\vecq{}$\,$\|$\,[110] and [111]. We can obtain a more detailed view at the directional dependence by constructing the difference spectra I$_{\vecq{}\,\|\,[100]}$\,-\,I$_{\vecq{}\,\|\,[110]}$ that is displayed as dichroism in the bottom panel(b) of Fig.\,\ref{Fig04} (black dots).

The Ce\,N$_{4,5}$ NIXS data are simulated by calculating the $4d^{10}4f^1$\,$\rightarrow$\,$4d^{9}4f^2$ transition using the full multiplet code \textsl{Quanty}\cite{Haverkort2016} which includes Coulomb as well as spin-orbit interactions. A Gaussian and a Lorentzian broadening of FWHM\,=\,0.7\,eV and 0.4\,eV, respectively, are used to account for the instrumental resolution and life time effects. The atomic Hartree-Fock values were adjusted via the peak positions, resulting in reductions of 30\,\% and 22\,\% for the 4$f$-4$f$ and 4$d$-4$f$ Coulomb interactions, respectively. The reduction accounts for configuration interaction effects not included in the Hartree-Fock scheme \cite{Tanaka1994}. A momentum transfer of $|\vecq{}|$\,=\,9.2\,\AA$^{-1}$ has been used for the simulations (and not the experimental value of 9.6\,$\pm$\,0.1)\,\AA$^{-1}$) so that the experimental peak ratio of the two main features around 108 and 110\,eV is reproduced best. This fine tuning optimizes the multipole contributions to the scattering functions to mimic for a minor adjustment of the calculated radial wave functions of the Ce$^{3+}$ \textsl{atomic} wave function (see e.g.\ Ref. \cite{WillersPRL109}).

We now compare the measured spectra and the dichroism therein with the simulations for the two possible scenarios, namely one with the $\Gamma_7$ doublet as ground state and the other with the $\Gamma_8$ quartet. The results are plotted in Fig.\,\ref{Fig04}\,(a). The $\Gamma_8$ quartet scenario reproduces in great detail the experimental spectra for all three $\vecq{}$ directions. Actually, the match is excellent. In contrast, the simulation based on the $\Gamma_7$ doublet exhibits large discrepancies with respect to the experiment: the intensities of several features in the spectra are not correct. To make the difference between the two scenarios even more contrasting, we compare the experimental and calculated dichroic spectra, i.e.\ I$_{\vecq{}\,\|\,[100]}$\,-\,I$_{\vecq{}\,\|\,[110]}$, as displayed in bottom panel(b). There is an excellent match for the $\Gamma_8$ quartet ground state scenario but a large mismatch for the $\Gamma_7$ doublet. From these comparisons we can unambiguously conclude that the $\Gamma_8$ quartet forms the ground state in CeB$_6$.

In addition, we have taken spectra at T\,=\,295\,K. The spectra look very similar to the low temperature data but the dichroism is reduced by about 20\%, see bottom panel\,(c) of Fig.\,\ref{Fig04}. This reduction in the dichroism is fully consistent with a partial population of the excited $\Gamma_7$ state at 46\,meV. A simulation in which the Boltzmann weighted contributions of the $\Gamma_8$ and $\Gamma_7$ states are taken into account is represented by the magenta line in panel\,(c) of Fig.\,\ref{Fig04}. The excellent agreement provides yet another evidence for the thorough understanding we have obtained using NIXS on the Ce\,$4f$ symmetry and crystal-electric field effects in CeB$_6$.

\section{Summary}

Using Ce\,N$_{4,5}$ non-resonant inelastic x-ray scattering (NIXS) we were able to establish that the ground state symmetry of the cubic hidden order compound CeB$_6$ is the $\Gamma_8$ quartet. The high signal to background ratio of the N$_{4,5}$ NIXS signal indicates that this bulk sensitive and element specific spectroscopic technique is a powerful method to study the local electronic structure of the rare-earth ions in rare-earth borides. With NIXS probing directly the charge distribution of the $4f$ electrons, it complements nicely neutron scattering based techniques which provide direct information on the spin distribution.

\acknowledgments
We thank P.~Thalmeier for valuable discussions. We also thank C.~Becker and T.~Mende from MPI-CPfS, and F.-U.~Dill, S.~Mayer, and other members from beamline P01 for their skillful technical support, C.J.~Sahle and M.~Harder for their valuable contribution to the data processing. Part of this research was carried out at the light source PETRA III at DESY, a member of the Helmholtz Association (HGF). K.C., M.S. and A.S. benefited from the financial support of the Deutsche Forschungsgemeinschaft (DFG) under projects SE 1441/1-2 and SE 1441/1-3. 
 
%


\end{document}